\documentclass[letterpaper]{article}
\usepackage{isea}
\usepackage[pdftex]{graphicx}
\usepackage{times}
\usepackage{helvet}
\usepackage{courier}
\usepackage{caption} 
\usepackage{cuted}
\usepackage[numbers]{natbib}

\title{Can Code Outlove Blood? An LLM-based VR Experience to Prompt Reflection on Parental Verbal Abuse}

\author{
    Jiaying Fu\thanks{These authors contributed equllly to the work.}, Jialin Gu\footnotemark[1], Tianyue Gong\footnotemark[1], Tiange Zhou\thanks{Corresponding author} \\
    School of Future Design, Beijing Normal University \\
    Zhuhai, China \\
    fujiaying@gmail.com, amo973571606@gmail.com, tianyuegong1203@163.com, tiangezhoumusic@gmail.com
}

\begin{document}
\maketitle
\begin{strip}
    \vspace{-5.5em}

    \centering
    \includegraphics[width=0.7\linewidth]{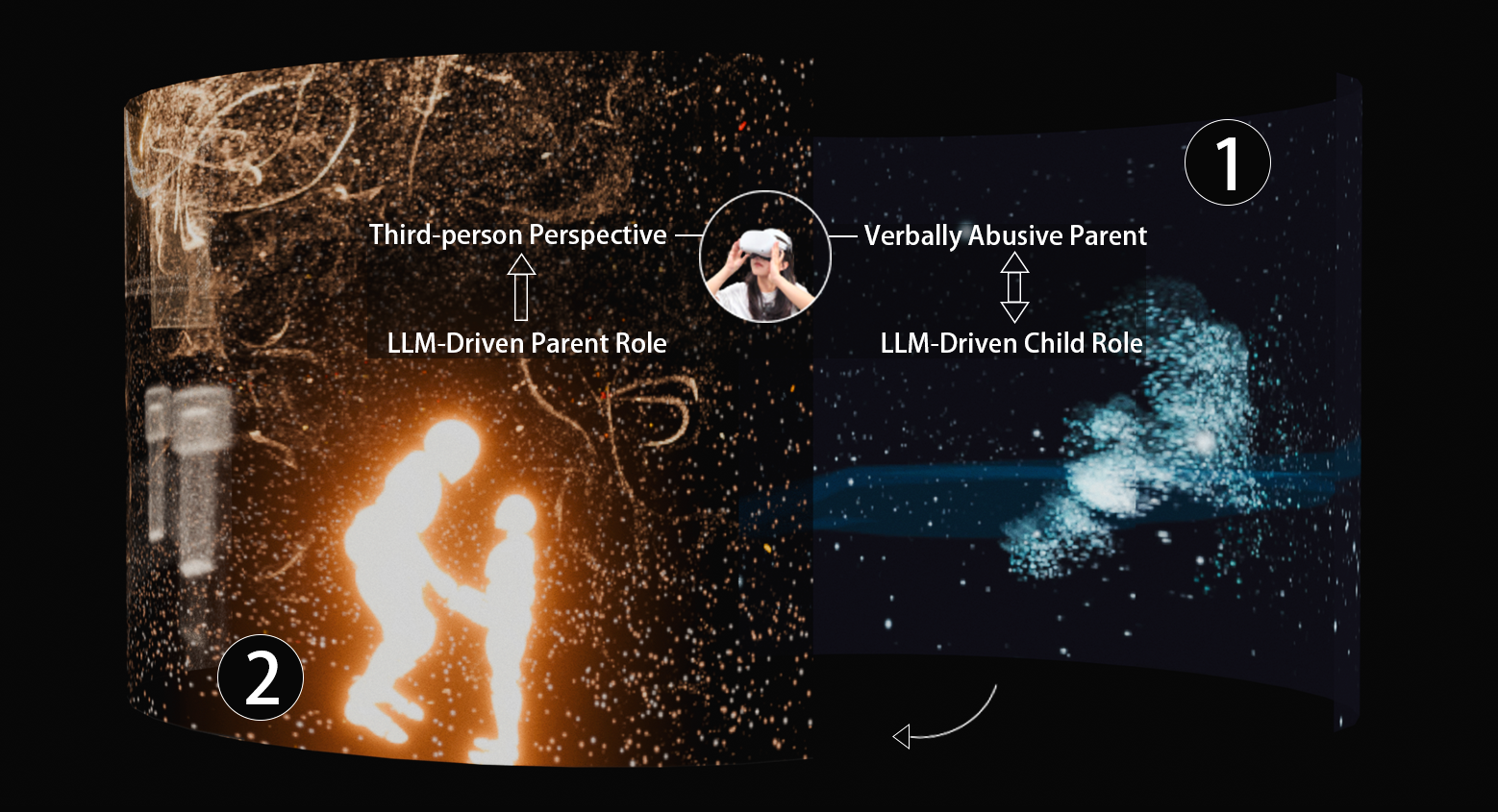}
    \captionof{figure}{A dual-phase VR system integrating  LLM-driven characters to explore parental verbal abuse. }
    \label{fig:enter-label}
\end{strip}

\begin{abstract}
% 摘要内容放在这里
Parental verbal abuse leaves lasting emotional impacts, yet current therapeutic approaches often lack immersive self-reflection opportunities. To address this, we developed a VR experience powered by LLMs to foster reflection on parental verbal abuse. Participants with relevant experiences engage in a dual-phase VR experience: first assuming the role of a verbally abusive parent, interacting with an LLM portraying a child, then observing the LLM reframing abusive dialogue into warm, supportive expressions as a nurturing parent. A qualitative study with 12 participants showed that the experience encourages reflection on their past experiences and fosters supportive emotions. However, these effects vary with participants' personal histories, emphasizing the need for greater personalization in AI-driven emotional support. This study explores the use of LLMs in immersive environment to promote emotional reflection, offering insights into the design of AI-driven emotional support systems.

%超字数的旧版 
%Parental verbal abuse leaves lasting emotional impacts, yet current therapeutic approaches often lack immersive self-reflection opportunities. To address this, we developed a VR experience powered by LLMs to foster reflection on parental verbal abuse. Participants with relevant experiences engage in a dual-phase VR experience: in the first phase, they assume the role of their verbally abusive parent, interacting with an LLM portraying a child; in the second phase, they observe the interaction from a third-person perspective, where the LLM acts as a nurturing parent, translating the previous abusive dialogue into warm, loving expressions. A qualitative experiment with 12 participants demonstrated that the experience encourages reflection on parental verbal abuse and fosters supportive emotions. However, these effects vary with participants' personal histories, emphasizing the need for greater personalization in AI-driven emotional support. This study explores the use of LLMs in immersive VR to promote emotional reflection, offering insights into the design of AI-driven emotional support systems.
\end{abstract}

%10/02暂定
VR-based Emotional Reflection, Parental Verbal Abuse, LLM-Driven Role-Playing
% The title ``Keywords'' should be 10 point, bold type, centered at the beginning of the left column. Using 9 point, justified, regular type, write up to ten keywords that highlight the main areas of your essay's content. 

\section{Introduction}

Parental verbal abuse leaves lasting emotional scars, influencing an individual's ability to regulate emotions and maintain healthy relationships. Characterized by critical, manipulative, or neglectful language, these interactions can deeply impact self-esteem, psychological well-being, and social functioning~\cite{danese2009adverse, ney1987does}. Research highlights that the effects of such abuse often persist into adulthood, intensifying vulnerabilities and hindering emotional resilience~\cite{lawson2013complex}. Despite growing recognition of its profound consequences, existing therapeutic approaches frequently fall short in providing immersive, reflective opportunities for individuals to process these experiences fully.

Traditional therapies such as cognitive-behavioral therapy (CBT) and narrative interventions have proven effective in guiding individuals to reinterpret their experiences and develop coping mechanisms~\cite{zayfert2019cognitive}. Beyond these, group-based methods like psychodrama and expressive arts therapy offer dynamic and participatory approaches by encouraging individuals to reenact family dynamics or externalize their emotions through creative expression~\cite{kellermann1992focus, malchiodi2011handbook}. While these methods have their strengths, group settings can create pressure, making some individuals uncomfortable with open expression~\cite{kellermann1999ethical}.

Advances in technology, particularly virtual reality (VR) and large language models (LLMs), suggest that technology might indeed offer new avenues to address these gaps. VR enables users to immerse themselves in controlled environments, facilitating engagement with emotionally charged scenarios~\cite{slater2016enhancing}. LLMs, on the other hand, generate dynamic and emotionally resonant dialogues, adapting to user input in real time. Together, these technologies offer immersive, reflective experiences, reducing discomfort from group settings and addressing accessibility challenges in traditional therapies.

However, the integration of such tools prompts deeper questions: Can artificial systems built on computational models foster the emotional depth often associated with human relationships? Can technologies like VR and LLMs, unburdened by biological ties, provide the understanding and care that familial bonds sometimes fail to offer? 
These reflective considerations form the foundation of this research, guiding the following questions:

\begin{itemize}
  \item How can a VR experience with LLMs be designed to engage participants with a history of parental verbal abuse and encourage reflection on their experiences?
  \item What emotional responses and reflections do participants report following their engagement with the VR-LLM experience?
\end{itemize}

\section{Background}
% Templates that implement these instructions can be retrieved at  {\small \tt http://isea2025.isea-international.org/}

\subsection{Parental Verbal Abuse and Emotional Well-Being}

Research indicates that experiencing parental verbal attacks during childhood is associated with increased levels of depression, anxiety, dissociation, and substance use in adulthood ~\cite{POLCARI201491}, and is more likely to develop a negative self-schema~\cite{GIBB2002223}. Besides the direct emotional harm, the cumulative effects of verbal abuse can also damage self-esteem, impede emotional regulation, and form a cycle of interpersonal conflicts that persists into adulthood~\cite{teicher2006sticks}. Those who have suffered from parental verbal abuse are more than three times as likely to experience borderline, narcissistic, obsessive-compulsive, and paranoid personality disorders during adolescence or early adulthood~\cite{johnson2001childhood}.

Conventional therapeutic approaches, such as CBT and family systems therapy, have shown efficacy in helping individuals process the effects of verbal abuse~\cite{cohen2015trauma}. However, these methods often rely on retrospective discussions in clinical settings, which may lack the immediacy and immersion necessary for deeply felt emotional reflection. Techniques like role-play and narrative therapy attempt to engage participants more actively, yet they are limited by the constraints of verbal recounting and the absence of experiential depth~\cite{fletcher2018more}.

%White1990NarrativeMT

To address this need, approaches like psychodrama have been developed to allow participants to reproduce family interactions within a controllable environment. Role reversal stimulates participants to adopt the perspectives of those with whom they have disputes or family issues, thereby enhancing their awareness of their own feelings and emotions.~\cite{chesner2019one-to-one-pshchodrama}. Through participating in the simulated family dynamics, individuals can gain an in-depth understanding of their own relationship patterns while maintaining an emotional distance.

Building on these methodologies, technology-driven tools like VR and LLMs provide new opportunities for creating immersive interventions. VR's ability to elicit emotional responses~\cite{felnhofer2015virtual} and LLMs' capacity for natural dialogue offer a promising direction for addressing the challenges of parental verbal abuse.

\subsection{LLM Intervention in Family Studies}

The theory and practice of parental verbal abuse is not a new thing. However, with the development of LLMs, different studies on the involvement of LLM in family relationships and mental health have gradually emerged in the field of HCI.

In fields of HCI research in family studies, Francisco Perella-Holfeld~\cite{10.1145/3678556} conducted research on parents and educators regarding children's use of AI robots in family and school environments. There are also studies on education. For instance, Hui-Ru Ho designed an educational robot to promote parent-child conversations about mathematics~\cite{10.1145/3585088.3589358},  and Di Liu's team's development of a new therapy through story creation and generative AI to foster family communication, supporting the needs of children, parents, and therapists~\cite{10.1145/3613904.3642852}. A few studies on the healing effect of LLM have emerged in the field of mental disease. For instance, Yongfu Wang proposed an LLM-based chat platform for children with autism as an early intervention tool and a bridge for parents to understand their children~\cite{10.1145/3675094.3678476}. He also discussed the advantages and disadvantages of using LLM for the diagnosis and personalized training of children with autism~\cite{10.1145/3675094.3677573}. Additionally, Yilin Tang focused on improving emotional recognition and expression abilities for children with high-function autism~\cite{10.1145/3613904.3642899}.

In the field of mental health, Harsh Kumar's team proposed that LLM chatbots have the potential to provide mental health support for computer science students~\cite{10.1145/3545947.3576285} in 2023. Further, in later 2023, Nima Zargham~\cite{10.1145/3571884.3603761} proposed that LLM with anthropomorphic traits can better enhance user experience and engagement. He constructed a humorous agent and discussed that LLM conversation agents are rapidly advancing in terms of capabilities and human portraits—both aiming to enhance user experience and engagement. Many studies have attempted to construct agents to deal with human emotions. For example, Abeer Alessa constructed an emotional companion for the elderly to help reduce loneliness and social isolation~\cite{10.1145/3594806.3596572}.

\section{Design}

To address our research question, we set the following design goals: to simulate family verbal abuse communication using character-specific LLMs; to implement a two-stage interaction based on psychodrama theory; and to design visual elements that enhance immersion.

\subsection{Overview of the VR Experience}
%In the virtual world, the audience is immersed in the virtual scene by wearing VR glasses. The whole experience consists of two distinct virtual scenes in which the audience will experience a role transition. Our design is based on the role reversal technique in psychodrama mentioned by PF Kellermann \cite{kellermann1994role}, which is in line with his view that role reversal is an effective therapeutic role playing tool.
The entire experience consists of two distinct virtual environments, where the audience undergoes role transitions between the scenes. Our design goal is similar to the role reversal technique mentioned by Kellermann in psychodrama~\cite{kellermann1994role}. We aim to promote emotional reflection through the changes in the audience's roles within the two scenes.

The first virtual scene presents a sad and gloomy environment, where the audience takes on the role of parents who pass on abusive words to their children. 
As shown in Figure 2, upon entering the scene, the audience is required to communicate hurtful words to the child through voice input according to a preset script. After that, the LLM-driven child responds to what the audience says, and the conversation continues for 3-4 rounds to deepen the interactive experience. This design aims to allow the audience to re-examine their own experiences through the perspective of the "perpetrator", thereby triggering reflection and emotional awareness.

\begin{figure}[h!]
    \centering
    \includegraphics[width=1\linewidth]{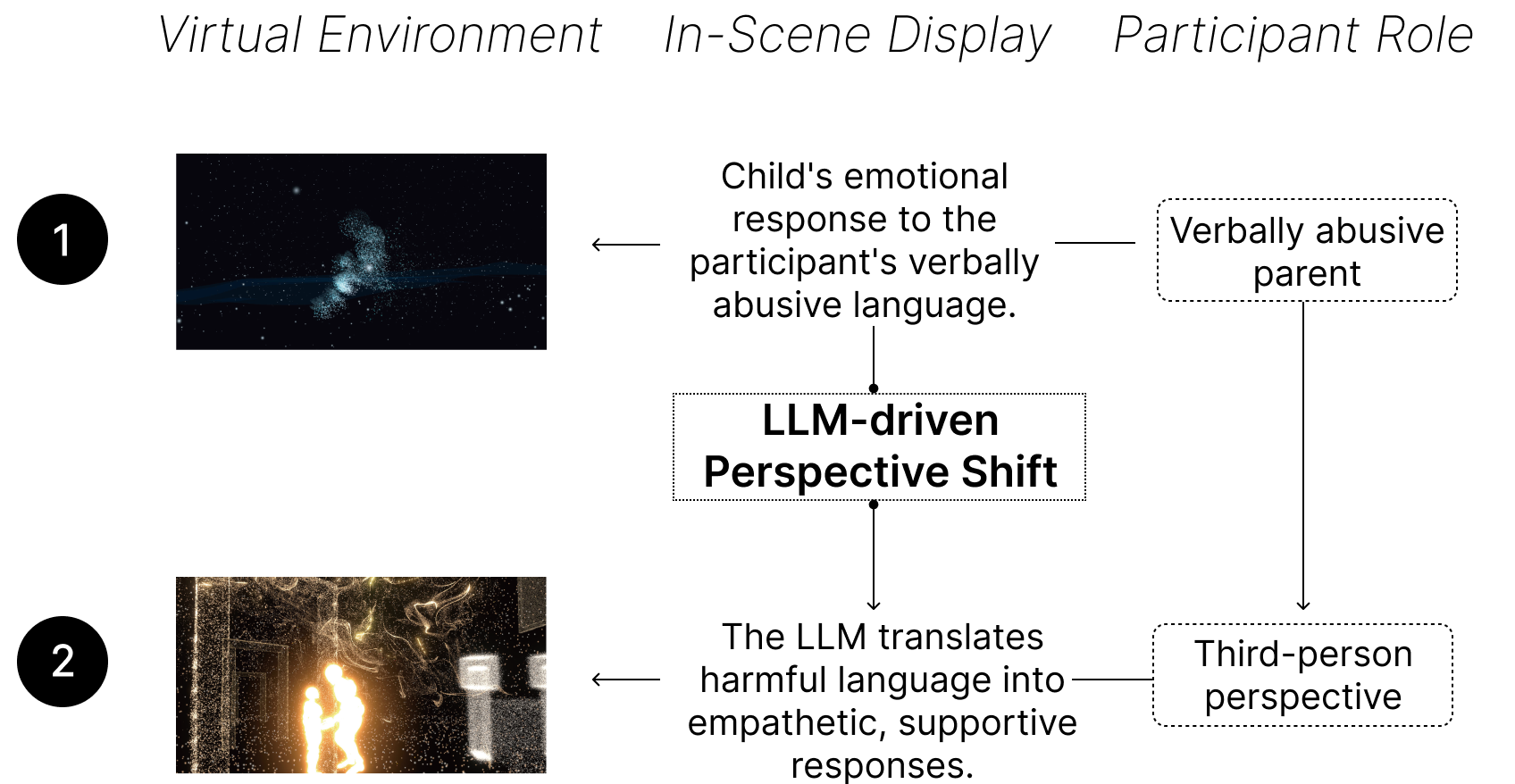}
    \caption{Dual-phase VR experience with LLM-driven perspective shift.}
    \label{structure}
\end{figure}

With the words of the child leaving in anger and the sound of the door breaking out, the first scene ends, and the audience seamlessly transitions to the second scene - a family full of warm atmosphere. In this scene, the audience's perspective shifts to a third-person point of view, becoming an observer of a family with healthy communication patterns (see Figure \ref{structure}). The mirroring technique in psychodrama therapy, as mentioned by Chesner~\cite{chesner2019one-to-one-pshchodrama}, is used here, allowing participants to observe their own scene from an external angle and gain a broader, more authentic perspective through third-party observation. At the same time, the LLM-driven mom character skillfully reinterprets the hurtful language in the first scene to communicate with the child in a warmer, more inclusive and harmless way. Yaffe noted that~\cite{yaffe2023systematic}, in parenting, mothers are generally more accepting, responsive, and supportive compared to fathers. Therefore, in our design, we chose the mother as the language translator to ensure that the emotional tone of the translated language aligns more closely with the audience's psychological expectations.

\subsection{Scene and Character Design}
In the visual scene design of this study, we use Unity Engine to make a semi-concrete performance of the scene, and introduce the particle effect of Visual Effect Graph into the design to give the scene more dynamic expression and emotional tension.

\begin{figure}[th!]
    \centering
    \includegraphics[width=85mm]{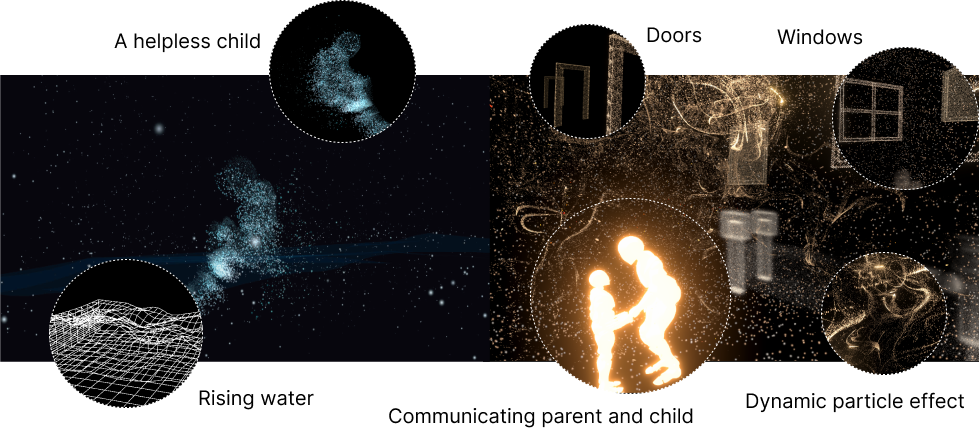}
    \caption{Visual elements of the VR experience.}
    \label{visual}
\end{figure}

Based on the theory proposed by Felnhofer et al.~\cite{felnhofer2015virtual} that emotions can be stimulated by adjusting virtual environment parameters (such as lighting, weather, etc.), we set the main color of Scene 1 as cold and dim. In the background design of the scene, referring to Chirico et al.'s research on the ability of certain visual elements to induce emotions~\cite{chirico2018designing}, we adopted deep tones and rising waters to strengthen the depressed atmosphere and metaphorically symbolize the sense of psychological pressure and emotional isolation experienced by children in negative language environments. In terms of character design, drawing on the theory of Barbara Pease and Allan Pease~\cite{pease2008definitive}, which suggests that individuals with nervous, negative, or defensive attitudes often adopt closed body language, such as tightly folding their arms across their chest, we depicted the child with a posture of sitting with their head down and knees clasped (see Figure \ref{visual}). This is designed to convey the character's sense of depression and symbolize the child's helplessness and self-protective instincts when facing challenges within a family environment. Besides, we further applied the effect of flowing and gradually dissipating particles, forming a visual atmosphere on the verge of fragmentation to enhance the expression of this emotion. With the audience constantly outputting negative language to the children in the scene, the water in the painting will continue to rise, gradually drowning the child, and the particles forming the child will gradually dissipate and finally completely integrate with the water.

\begin{figure*}[ht!]
    \centering
    \includegraphics[width=160mm]{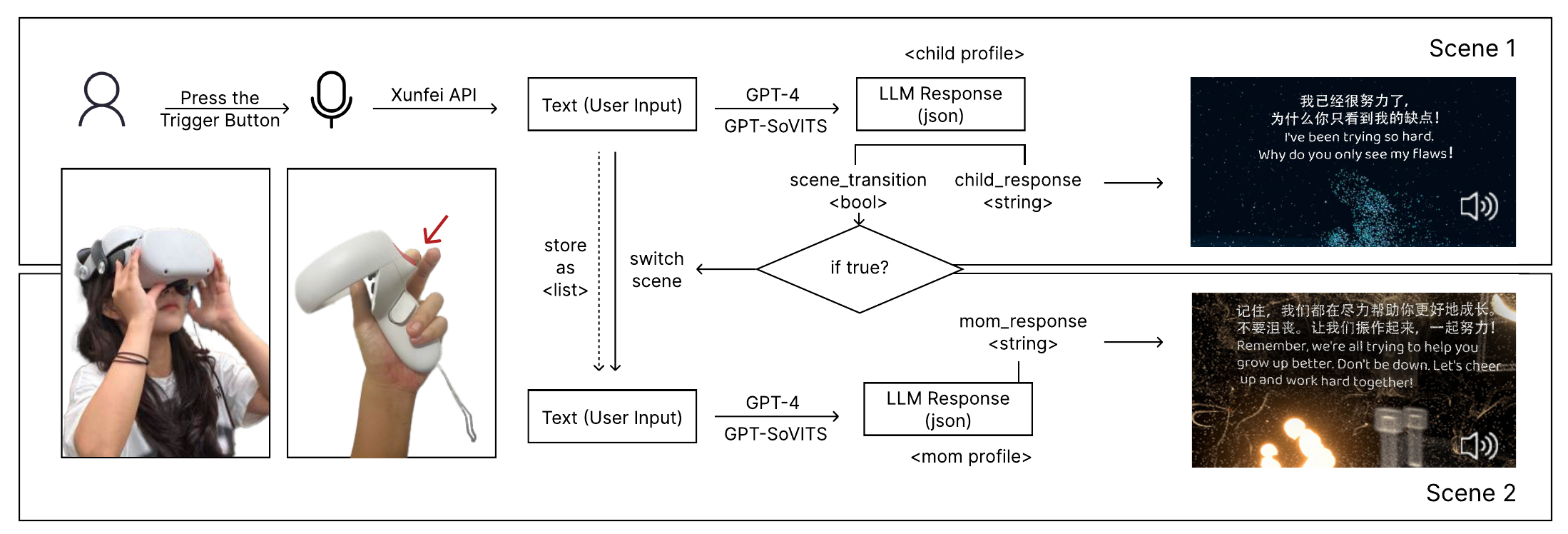}
    \caption{VR system architecture depicting the dual-scene interaction.}
    \label{system}
\end{figure*}

% \begin{figure}
%     \centering
%     \includegraphics[width=80mm]{ISEA/scene2.jpg}
%     \caption{The water in the painting will constantly rise and submerge the child, and the particles that make up the child will gradually dissipate}
%     \label{fig:enter-label}
% \end{figure}

In transitioning between the two scenes, the narrative is designed so that the child exclaims, "Stop it, I'm leaving!" This is immediately followed by the sound of a heavy door closing, signaling the switch to the second phase in the VR experience. This design aims to symbolize the child's emotional frustration and sense of desperation, while also ensuring a seamless narrative transition that maintains the coherence of the overall experience.

In Scene 2, we build a warm and safe family space to create a harmonious and comforting atmosphere. The whole scene is set in a simple bedroom, which includes warm and familiar domestic elements such as doors, windows, and a bed that symbolize "home" (see Figure \ref{visual}). The room is bright and warm, filling the entire space with tranquility and comfort~\cite{felnhofer2015virtual}.
Based on the theory of Barbara Pease and Allan Pease~\cite{pease2008definitive}, we design the parents in the scene to gently bend down and talk with their child, symbolizing an equal communication dynamic (see Figure \ref{visual}). Both parents and the child emit a warm halo, a faint glow that slowly changes over time, symbolizing the flow and continuation of intimate relationships.

%In Scene 2, we build a warm and safe family space to create a warm and harmonious atmosphere. The whole scene is set in a simple bedroom, which includes warm and familiar domestic elements such as doors, windows and bed that symbolize "home". The room was bright and warm, makes the whole space brimming with tranquility and comfort~\cite{felnhofer2015virtual}. Based on the theory of Barbara Pease and Allan Pease~\cite{pease2008definitive}, we designed the parents in the scene to gently bend down and talk with their children, symbolizing that the two people are placed in an equal communication position. Both parents and children emit a warm halo, a faint halo that slowly changes over time, as if symbolizing the flow and continuation of intimate relationships.

This warm atmosphere is further enhanced by the dynamic particle effect, which is based on Pinilla et al.'s research on the use of abstract elements to trigger specific emotions in users~\cite{pinilla2021visual}.
Small particles float in the air like ribbons, gently permeating the sky and slowly spreading throughout the room, symbolizing the extension of emotions and the intangible infectious power of family warmth. Over time, these particles gradually spread throughout the entire scene, creating a serene and inclusive emotional atmosphere.
% \begin{figure}
%     \centering
%     \includegraphics[width=80mm]{ISEA/scene1.jpg}
%     \caption{Small particles float in the air like ribbons, gently permeating the space. Parents gently bend down to talk to their children.}
%     \label{fig:enter-label}
% \end{figure}

\subsection{System Implementation}
The system, developed on Unity 2022.3.21f1 and deployed on Oculus Quest 2, provides an immersive VR environment where users interact with two LLM-driven characters across two scenes. In Scene 1, users initiate voice interaction by pressing the controller's Trigger button, which activates the microphone to capture audio input. This audio is processed by the Xunfei API to convert it into text, which is then sent to GPT-4 to generate responses from the perspective of a "child" character (see Figure \ref{system}). The generated response is synthesized into voice via the GPT-SoVITS Fast API, aligning with the child's profile, using voice models trained on authorized audio from the author and their mother to ensure authenticity.

In Scene 1, prompt engineering guides GPT-4 to role-play as a 16-year-old who regularly faces parental verbal criticism related to appearance, academics, and personal habits. The prompt directs GPT-4 to convey escalating emotions of frustration and helplessness, intensifying over 3–4 dialogue exchanges until an emotional breaking point is reached. The system transitions to Scene 2 when the \textit{scene\_transition} field in GPT-4's response is set to \textit{true}, indicating a culmination in the child's emotional breakdown. Each user input in Scene 1 is stored to facilitate continued reflection in Scene 2.
In Scene 2, stored inputs from Scene 1 are sequentially sent to GPT-4, now configured to respond as a "mom" character. For each input, GPT-4 generates responses from the mom's perspective, which are then voiced through the GPT-SoVITS Fast API, producing a warm, supportive output characteristic of the mom profile (see Figure \ref{system}). The prompts in Scene 2 direct GPT-4 to rephrase the child's frustrations in an empathetic and constructive tone, simulating a nurturing mother who addresses concerns with compassion~\cite{afifah2016keep}. This setup aims to promote reflection by offering responses that avoid emotional harm, maintain parental intent, and encourage constructive feedback, with each response formatted in JSON under the \textit{mom\_response} field.

Each scene displays the LLM's responses through both audio and a text UI component within the VR environment. By employing detailed prompt engineering that includes specific character settings and multimedia design, the system achieves a character-specific interactive experience with seamless transitions between the two roles.

\section{Methods}
% Based on the real cases of experiencing parental verbal abuse, we transformed these language texts using the GenAI system and elaborately designed a set of immersive VR interactive devices, encompassing the layout of scenarios and dialogues. After recording the experience and viewing process, interviews were conducted with the participants respectively, and qualitative analysis was carried out on the results. 
\subsection{Participants}
This study recruited 12 participants from China, including 4 males and 8 females aged 18 to 34. Participants were required to have the experience of parental verbal abuse from parents in daily life. 

In addition, participants needed to pass the self-rating depression scale (SDS)~\cite{Zung1965} to evaluate their mental health, and those with severe depression (Y
\textgreater 0.6) were excluded by SDS in consideration of experimental ethics. All participant information was strictly confidential, and data were processed anonymously. During the experiment, participants could withdraw at any time.

\subsection{Procedure}
\begin{figure}[h!]
    \centering
    \includegraphics[width=1\linewidth]{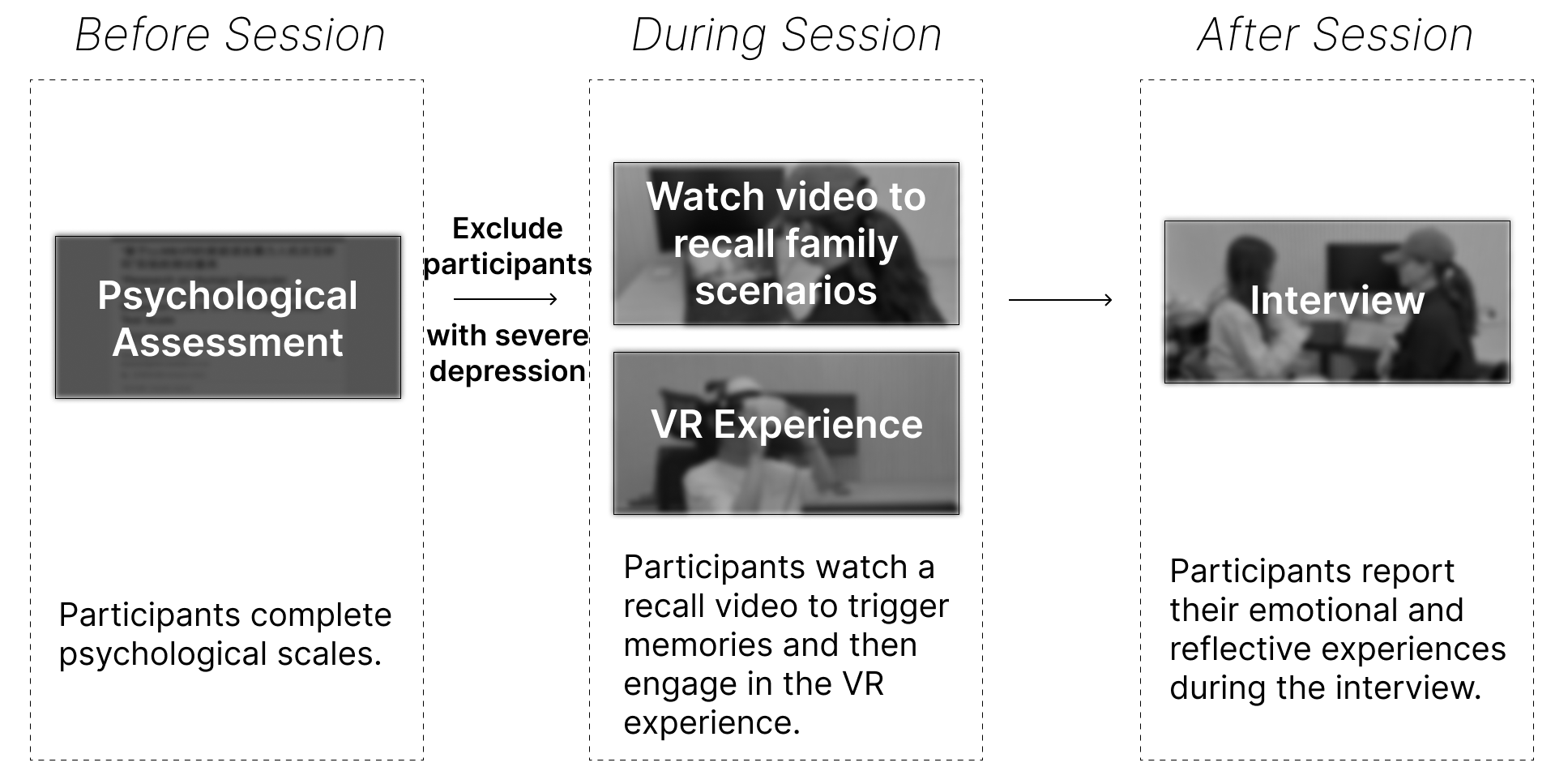}
    \caption{Overview of the experimental procedure.}
    \label{methods}
\end{figure}

For the pre-experimental preparation, we introduced the purpose, process, data collection method and ethical considerations of the study to participants to ensure full understanding and signed the informed consent.
We played a segment from a TV drama depicting a mother verbally abusing her daughter, sourced from popular clips on the widely known Chinese social media platform Xiaohongshu to help participants recall and bring into the scenes of familial verbal abuse.
Then, participants experienced a dual-stage VR experience and participated in interviews afterwards (see Figure \ref{methods}).

Our interview aimed to gather insights on participants' experiences and reflections during the dual-phase VR interaction. Specifically, we sought to understand: i) how the language used in participants' role-playing in Scene 1 related to their past experiences, ii) how participants perceived the LLM's responses in Scene 1, including their emotional triggers, realism, and whether the responses prompted further dialogue or evoked memories of their past selves, and iii) participants' overall emotional reactions to Scene 1, with a focus on identifying specific VR design elements that influenced these reactions. For Scene 2, we explored: iv) participants' perceptions of the LLM's ability to reinterpret prior dialogue, v) evaluations of the communication style and realism, and vi) what feelings and perceptions this interaction evoked for them. Additionally, we sought feedback on vii) participants' overall reflections on Scene 2, particularly how VR elements shaped their engagement and emotional responses. Beyond the VR interaction, we investigated viii) participants' prior experiences discussing psychological or emotional issues with LLMs, ix) their perspectives on AI-driven emotional support tools, and x) any connections or differences they perceived between the VR experience and other similar applications.

\subsection{Analysis}

In this experiment, we conducted a qualitative analysis of interview data using open coding and thematic analysis. Participants' feedback was categorized into key themes: "recollection of past events" captured responses about memories evoked in Scene 1, "emotional impact" reflected feelings such as unease or comfort in Scene 2, "LLM response effectiveness" encompassed evaluations of the dialogue's realism and its ability to elicit emotions, and "environmental design influence" highlighted feedback on VR elements like color tones and particle effects. These themes allowed us to analyze how participants engaged with the VR-LLM experience and its emotional effects.

% The interview content will also be recorded in written form and analyzed one by one:

% 1.How is the influence of role-playing on the perception of past experiences? - How to design a virtual reality experience with a large language model to enable participants to engage in parental verbal abuse and encourage reflection?

% 2.How to evaluate triggering effect and authenticity of LLM responses? - How does the two-phase virtual reality experience affect participants' emotional reflection and understanding of parental verbal abuse? 

% 1. The impact of psychodrama method on the view of past experience? ——Participants' feelings about reproducing the experience of domestic language violence in VR scenes

% 2. Will LLM's response inspire you to say further hurtful words? ——The impact of LLM (large language model) dialogue on participants' cognition and emotions

% 3. Is the gentle parent LLM in the second scene real and is it a better family model? ——Participants' views on discussing psychological problems with LLM

\section{Qualitative Findings}
\textbf{The VR-LLM design engages participants by simulating parental and child roles, enhancing emotional reflection on past experiences.}  All participants apart from P4 stated that in the first scene, they reproduced some or all of the scenes that occurred at home and the experience was very similar. They thought of their past selves and substituted them into the LLM child role. P1 evaluated the dialogue as "like returning to childhood and watching the past at home".

The interaction with the child played by the LLM promoted the role-playing of the participants. At the beginning of the interaction, some participants (P5, P8, P9, P10) expressed difficulty in articulating abusive language in their role as a parent. However, the responses from the LLM child gradually elicited emotional expressions (P5, P6, P11, P12), deepening the parental role-play. As P11 mentioned: "When the LLM mentioned a specific word, it triggered my memory, reminding me of what my mother said at that time."
%After ending the conversation with LLM, all participants apart from P3 find it hard to get rid of the child's perspective. Although many participants (P4, P6, P7, P8, P9, P10) can empathize more with parents, they consider the words inappropriate and overly hurtful.

The interaction in role-playing stimulated the participants' reflection of their past experiences. Participants P1, P3, P6-P11 think LLM responses are relatively realistic. The rebellious language style (P1, P3, P6, P10), the self-expression communication mode (P7, P8, P11), and the act of slamming the door and leaving (P4, P9) enable participants to see past shadows. For example, P3 thinks naive and rebellious language like "Why can't you consider my feelings?" and acts like slamming the door happened in personal experiences. 
In addition to noticing similarities with the responses of the child portrayed by the LLM, some participants also highlighted differences between their own reactions and those of the LLM in the same scenarios, which led to varied experiences. For instance, P8 expounded that, in contrast to the highly expressive responses of the LLM child, maintaining silence and adopting avoidance behaviors were more likely to be his strategies in response to parental censures in actual situations. P9 stated that she would employ even more extreme words and intonations than those of the LLM to vent her emotions, such as mentioning suicide. P11 perceived that the postures of the child exhibited an introverted and defensive nature, whereas the language of the LLM was rather rebellious, which led to a sense of fragmentation.

The elements within the scene inspired diverse feelings among the participants, and these feelings were in line with the overall experience. Many participants (P1, P2, P4, P6, P10) described Scene 1 as somber and emotionally heavy. Some of them held the view that it was the cold and depressing environmental colors that intensified these feelings (P1, P6). Participant P10 thought that the postures of characters curling up and the direction of facing sideways towards the participants enhanced the experience of ineffective communication. However, some participants, including P3 and P5, described the scene as gentle and calm, aligning with their impression of the LLM child's responses. P5 noted, "The tone of the LLM was much calmer than my own in reality."

\textbf{The dual-phase VR experience prompts reflective and supportive feelings, yet evokes varied emotional responses shaped by personal histories.}
Most participants perceived that in Scene 2, the LLM had effectively translated their previously expressed abusive language into a gentler, solution-oriented style (P1, P2, P3, P4, P6, P7, P8, P9, P10, P11, P12). This rephrased language felt supportive and modeled healthier parent-child communication, which helped to soothe emotions (P1, P2, P3, P5, P6, P9, P10, P11). For some participants, the responses approached an idealized parental figure (P8, P9, P10), with P9 noting, "I wish my parents would speak to me this way." Others valued the LLM's consistent rationality during emotional conflict; as P7 commented, "During conflicts, people are driven by emotions, and that's when a consistently rational role like the LLM becomes valuable." Participants such as P6 and P9, who recalled past selves in Scene 1, found the supportive language in Scene 2 profoundly moving,  feeling finally understood.

The realism of the LLM's voice and gentle tone in Scene 2 enhanced authenticity for participants (P4, P6, P8), with P10 describing a slight fear at its human-like quality: "It sounds so human, yet it's a machine." The shift to warm color tones further created comfort (P1, P3, P5, P6, P7, P10, P11), while familiar home elements such as doors and windows (P4, P7, P8, P12), the image of two people embracing (P10, P11), and floating particles contributed to an inviting atmosphere (P6, P9). P5 described a nostalgic quality, as if the scene evoked deeply stored memories, while P6 felt the drifting particles symbolized growing closeness with family members, and P12 saw the open door as an invitation for communication.

Reactions were often tied to personal histories. For instance, P6 felt they were portraying a father unable to communicate well, while the mother in Scene 2 resembled a nurturing, real-life figure. P9 described the LLM-driven mother in Scene 2 as a "more emotionally stable" version of their own mother, enhancing the realism of the experience. However, some participants found the responses overly idealized and different from typical parental interactions (P3, P4, P7, P11, P12). P3, P4, P11, and P12 remarked that their own parents were more reserved and would not typically speak in this way, making the LLM's response style feel unfamiliar to them. Some likened the responses to textbook examples or the language of a psychologist (P4, P5, P7, P12), and P2 described it as observing another family's happiness from a distance. This idealized style occasionally created emotional distance, contrasting the LLM's responses with participants' lived realities. Nonetheless, many still viewed the LLM's responses as a valuable model for real-life communication (P1, P3, P7, P8, P9, P10).

In summary, while the VR experience fostered a supportive atmosphere, participants' emotional reactions varied, influenced by how closely the LLM's responses aligned with their personal histories.

\textbf{Participants view LLM as a potential tool for emotional support but highlight the need for personalized interactions in therapeutic settings.}
As mentioned above, participants generally felt that while the LLM had potential for emotional reflection, its responses resembled textbook examples or the overly formal language of a psychologist (P4, P5, P7, P12), leaving it unable to meet deeper emotional needs. Some participants (P4, P5, P12) pointed out that the LLM's response was not targeted and was more "rhetorical" and could not provide personalized support continuously. For example, P4 believes that although AI's comfort is effective at the beginning, it is difficult to support emotional needs in the long run due to lack of understanding of specific situations.

%Still, many participants expressed anticipation for the potential of AI emotional support system.In our study, P8 believes that although the LLM's response still lacks the flavor of life, if it can be closer to humans, it can achieve the desired support effect. P5 mentioned that she will indeed choose AI as the object to talk to when she encounter things that she cannot talk to others. ; P7 believes that AI's rationality can provide stable support when emotions are excited. P9 even hopes that this technology will help parents improve communication with their children.

Still, many participants expressed anticipation for the potential of the AI emotional support system. P8 said that although the LLM's responses lack the flavor of life in the experience, beyond that, it can shape the ideal image of parents that people have. P5 mentioned that she would indeed choose AI as the object to talk to when she encountered things she could not talk to others about. P7 believed that AI's rationality could provide stable support when emotions were excited. P9 even hoped that this technology would help parents improve communication with their children.

Participants positively evaluated their sense of immersion in this VR experience. They believed that compared to text-based LLM emotional support system, which primarily rely on text-based communication, our work provide a stronger sense of immersion through the integration of visual and voice elements (P10, P11). However, there is still room for improvement when it comes to scene realism. P8 mentioned that the lack of realistic objects in Scene 1 made the space feel empty. P9 believes the characters in Scene 1 should be more specific, as the current character design does not allow her to immediately connect with them. P12 indicates that the lack of familiarity affects the sense of substitution. These responses suggest that while LLM is recognized for emotional comfort, it still has potential for improvement in terms of personalization, persistence, and scene realism.

\section{Discussion}
\subsection{Design Implications}

This study integrates LLM-driven characters into a dual-phase VR experience, supporting users' emotional reflection in virtual parent-child interactions. Similar to Hu et al.~\cite{hu2024grow}, who used an "AI Buddy" to build children's mental resilience, our research also facilitates reflection through immersive, role-based interactions. However, we focus specifically on parental verbal abuse, using a role-switching approach to move between conflict and reconciliation, which deepens users' reflective experience across perspectives. While Lissak et al.~\cite{lissak2024colorful} applied tailored prompts to support LGBTQ+ youth with a therapist character, our study extends this by establishing a simulated parent-child relationship with the LLM, enhancing immersion and emotional reflection. Consistent with their conclusions, our findings also emphasize the importance of personalization in emotional support. Methodologically, this study uniquely combines a first-person conflicting" role with a third-person "observer" perspective for reflection. This dual-phase approach enables users to revisit their experiences from varied viewpoints, fostering reinterpretation of emotions within a supportive setting. Similar to Peck et al.'s~\cite{peck2013putting} work on fostering empathy through role-play, our multi-perspective design encourages users to engage with challenging emotions from varied viewpoints.

The design implications of this study suggest several directions for future emotional support systems. First, integrating emotional transitions and multi-stage interactions within VR could aid users in processing complex emotions, offering layered emotional support. Second, our focus on environmental design and multi-perspective prompt engineering could serve as a valuable framework for LLM-based support systems, enhancing emotional engagement and effectiveness. Finally, this study indicates that closely aligning LLM character settings with users' experiences is crucial for perceived authenticity. Future systems could benefit from embedding LLMs in immersive, multimedia environments that simulate realistic interpersonal scenarios, creating a more meaningful and engaging LLM-based emotional support experience.

\subsection{Limitation}

\textbf{Demographic Limitations in Participant Representation.} Participants in this study were limited to Chinese populations and lacked representation of a wide range of ethnic and cultural backgrounds. Although definitions of domestic verbal abuse are similar across ethnic groups~\cite{ferrari2002impact}, there are differences in the risk and frequency of domestic verbal abuse across ethnic or racial groups~\cite{luken2021On_racial_disparities_in_child_abuse_reports}. This difference results in a sample that does not fully reflect the true picture across all ethnic groups, limiting the generalizability and validity of the results. 

\textbf{Ethical considerations of LLM-driven VR experiences.} This study engaged participants in interactions with an LLM portraying a child character in a VR environment, requiring them to express critical or accusatory language. While this facilitated emotional reflection, it risked psychological discomfort. Research indicates that revisiting negative experiences can trigger distress, including PTSD symptoms~\cite{brewin1996dual}. Although no participants explicitly reported such effects in our study, the possibility of latent or delayed emotional reactions cannot be entirely dismissed.

Ethical concerns also arise from using LLMs in emotionally sensitive contexts~\cite{luxton2016ethical}, including biases in responses, pseudo-rational explanations, and varying effectiveness across diverse backgrounds~\cite{ntoutsi2020_llm_bias}. These considerations highlight the importance of further research to ensure these systems are both inclusive and psychologically safe.

\section{Conclusion}
This study presents a dual-phase VR experience powered by LLMs, designed to encourage reflection and emotional support regarding parental verbal abuse. In the first phase, participants role-play as verbally abusive parents interacting with an LLM portraying a child. The second phase offers a reflective perspective, where participants observe the LLM reframe abusive dialogues into nurturing expressions. Results from 12 participants show that the LLM-based experience fosters reflection on past abuse and promotes supportive feelings. However, emotional support is closely linked to participants' personal histories, emphasizing the need for greater personalization in AI-driven interactions. This research explores LLMs' role in emotional reflection within immersive experiences focused on family dynamics, with future work aiming to enhance personalization for improved engagement in emotional support systems.
\section{Acknowledgements}

This research has been supported by Beijing Normal University Research Fund (Grant No. 31220502509).

\bibliographystyle{isea}
\bibliography{main}

% \section{Author(s) Biography(ies)}
% The title ``Author(s) Biography(ies)'' should be 12 point, bold style, centered. Using 9 point, regular type, biographies should be no longer than 150-word count.

% \section{Questions?}

% For technical questions about Microsoft Word formatting please seek online tutorials. For other questions about your manuscript please contact: {\tt isea2025@nabi.or.kr}

\end{document}